# COMMENTARY:

# The evolving usefulness of the Test-Negative Design in studying risk factors for COVID-19 due to changes in testing policy.

Jan P Vandenbroucke[1,2,3], Elizabeth B Brickley[2], Christina M.J.E. Vandenbroucke-Grauls[4], Neil Pearce[2]

(1) Department of Clinical Epidemiology, Leiden University Medical Center, The Netherlands; (2) Departments of Medical Statistics, Non-communicable Disease Epidemiology and Infectious Disease Epidemiology, London School of Hygiene and Tropical Medicine, London, UK; (3) Department of Clinical Epidemiology, Aarhus University, Denmark; (4) Department of Medical Microbiology and Infection Prevention, Amsterdam UMC, The Netherlands

**Corresponding Author**:  Jan P Vandenbroucke, Leiden University Medical Center, Dept. Clinical Epidemiology, PO Box 9600, 2300 RC Leiden, The Netherlands, E-mail: J.P.Vandenbroucke@lumc.nl

**Running head**: Evolving usefulness of test-negative design COVID-19

The authors have no conflict of interest
No specific sources of financial support, other than salaries by affiliations

**Word count**: text: approximately 1369

# COMMENTARY:

## The evolving usefulness of the Test-Negative Design in studying risk factors for COVID-19 due to changes in testing policy.

In a recent paper in this journal,[1] we described the possibilities of using the Test-Negative Design (TND) as a fast and efficient tool for identifying local (e.g., attending a local concert) and general (e.g., working in the food service industry) risk factors for COVID-19. The basic idea of the TND is that those who seek testing for a particular disease may be motivated by a number of factors, such as the presence of symptoms, availability of transport, or access to affordable care. When not everyone in a population is tested, the persons who are tested will have gone through this 'selection' process, whether or not they test positive or negative. Thus, these selection factors are controlled for by selecting the test-positives as the cases and the test-negatives as the controls.

In the beginning of the COVID-19 pandemic, many countries prioritized people with symptoms for SARS-CoV-2 testing by reverse transcription polymerase chain reaction (RT-PCR) in order to quickly make diagnoses and institute isolation, contact tracing, and quarantine.  In this situation, the test-negative controls will have presented with similar symptoms, which may be caused by another respiratory pathogen (e.g., influenza virus), which may have some  transmission risk factors in common with SARS-CoV-2 (e.g., residing in congregate housing). Because of this potential challenge with using symptomatic test-negative controls, we advocated in our paper for the use of an additional control group representing the background population.

However, as the epidemic progressed, persons without symptoms were also increasingly being tested because, for example, of a recent contact with an infected person or as a precautionary measure to use certain public spaces or to participate in particular activities (e.g., flights). In the current paper, we focus on issues relating to the use of the test-negative design when an increasingly high proportion of the tested population includes persons without symptoms. It thus represents an extension of our previous work in which we assumed that almost all test-positives and test-negatives have symptoms. We continue to discuss these issues primarily in terms of RT-PCR tests; however, if a TND study is set up in a scenario with rapid tests (e.g., SARS-CoV-2 lateral flow assays), similar principles will usually apply.  However, owing to differences in test sensitivity, specificity and predictive values between different assays (PCR vs. rapid tests or lateral flow test) and because the use of a given assay may vary, based on the presence of symptoms, care should be taken to account for the type of assay used in differentiating the test-negative and test-positive persons – the principles of which were discussed in Appendix B of our previous paper.[1]

When more and more persons without symptoms are tested, but not everyone in a population is tested, there will still be selective pressures on who gets tested and who does not, and therefore there will remain a need for the TND. Thus, we wish to consider the application of the TND to the situation where not everyone in a population is being tested, but *those who do get tested are a mix of people with and without symptoms.* The selective pressures will be different for these two groups (e.g., the frequency of testing may be higher among those with symptoms than those without symptoms).

How should this be addressed in the analysis? The simplest approach is to record for each person tested whether the reason for testing is having symptoms or not, and control for this, either by doing a separate analysis by reason for testing, or adjust, depending on the type of question to which the analysis is intended. While some in the group with symptoms may have less severe symptoms than others (e.g., because some persons tend to approach health facilities more readily than others), this selective pressure will be similar for symptomatic test-positives and test-negatives,[2] and there is no need for control for severity of symptoms for reasons of validity. Although adjusting by severity has been advocated for reasons of non-collapsibility (i.e., in order to make inferences from the study sample more generalizable to the source population),[3] such problems of non-collapsibility are usually trivial.[4]

As testing becomes more affordable and more widely available, there will be heterogeneity in the reason for being tested, to a small degree in the group with symptoms, but to a much larger degree, in the group without symptoms.  Ideally therefore, the reason for seeking testing should be recorded and used in the analyses.

For example, for individuals tested without symptoms, two main reasons could be distinguished that would influence the selective pressures to be tested:

- testing because of the possibility of a close contact (e.g., with a family member, a workplace exposure, a warning signal by a phone app). Within this reason, it might be useful to differentiate according to the type of close contact. If the close contact is a family member, specific subquestions in a TND might elucidate family situations which lead to transmission more often (e.g., contact with grandparents or children of different ages). Further inquiry into the vaccine status of the person seeking testing may inform estimates of vaccine effectiveness against infection (and notably the likelihood of asymptomatic infection), while collecting additional data on the vaccine status of family contacts may shed light on vaccine effectiveness against infectiousness (e.g., comparing contact with vaccinated versus unvaccinated family members among unvaccinated persons). If the close contact is a co-worker or friend who tested positive, subquestions in a TND questionnaire may identify situations of increased risk, which may lead to some refinements in general precautions (e.g., frequency of outdoor versus indoor meetings).  If the contact is a warning by a phone app, this does not lead to inquiring about contacts with particular persons, but it leads to an inquiry about general risk situations of those testing positive vs. negative, e.g., type of occupation, having participated in mass events or indoor gatherings, use of face masks and keeping distances. The latter information is very useful to evaluate general precautionary measures. It might lead to studying the usefulness of phone apps by calculating how often people were referred by these apps, how often positive tests followed and modelling how much this may have helped to quell the epidemic.[5]

- testing as a precautionary measure; say, before visiting an elderly relative in a nursing home, or to attend certain activities like a concert when countries or regions are gradually relaxing distance rules. Here, testing is not based on symptoms, or on contacts, and the population being tested is therefore more comparable to a general population sample. It might be better to first stratify for these different reasons, before lumping them as a group.

The type of information that might be gleaned from comparing test-positives vs. test-negatives in this situation, might be comparable with the persons warned by a phone-app, since in both instances, a positive test will come as a surprise, and most likely no particular contact might be recalled.

It is important to realize that a TND among symptomatic persons provides information on the risk factors for symptomatic COVID-19, while our proposed extensions to persons without symptoms will yield information about risk factors for any infection with SARS-CoV-2. The risk factors for becoming infected are likely to be different from the risk factors for developing symptomatic disease. For example, increasing age may not be a strong risk factor for SARS-CoV-2 infection (on the contrary, younger persons may more often be infected due to more social contacts), but it is a strong risk factor for symptomatic COVID-19. The same may apply to other risk factors, such as sex and comorbidities. [ref multistep paper]. In other words, risk factors for developing disease could be thought of as primarily 'biological' depending on underlying health conditions, immune experience (e.g., via vaccination or prior infection), and infectious dose. However, risk factors for being infected could be considered largely 'social/behavioural' – in the sense of factors that increase the risk of exposure to infection (e.g., occupation, housing density, type of personality and personal activities). Comparing the findings of a TND in persons with symptoms versus the findings of a TND in persons without symptoms, while accounting for the reasons for seeking testing, could begin to differentiate between risk factors for infection and risk factors for disease after infection.